 \documentstyle[preprint,aps]{revtex}

\def\het{$^3\!$He }
\def\hef{$^4\!$He }

\begin{document}
\draft
\preprint{IC/94/1}

\title{    Variational Calculations for \het Impurities \\
                   on \hef Droplets }

\author{   A. Beli\'c$^a$, F. Dalfovo$^b$, S. Fantoni$^c$, and
                     S. Stringari$^b$ }

\address{$^a$ International Center for Theoretical Physics, \\
              P.O.Box 586, I-34100 Trieste, Italy }
\address{$^b$ Dipartimento di Fisica, Universit\`a di Trento, \\
              I-38050 Povo, Italy }
\address{$^c$ Interdisciplinary Laboratory, SISSA, \\ Via Beirut
		2, I-34014 Trieste, Italy }

\maketitle

\begin{abstract}
Variational Monte Carlo method is used to calculate ground
state properties of \hef droplets, containing 70, 112, 168,
240, 330, and 728 particles.  The resulting
particle and kinetic energy densities are used as an
input in the Feynman-Lekner theory for  \het impurities. The
kinetic energy density of \hef atoms and the energy of the
\het surface states are compared  with the results of previous
phenomenological calculations.
\end{abstract}

\pacs{67.40 }

\narrowtext

\section{Introduction}

Helium droplets have attracted some interest in recent years.
A major motivation is the fact that they represent a prototype of
finite size quantum systems, behaving as fluid of strongly
interacting structureless particles. Several
theoretical schemes have been developed for pure helium droplets
\cite{pan83,pan86,str87,cep89,wha90,kro90}. The comparison between
theory and experiments is still elusive, mainly because helium
droplets are so weakly bound objects that their experimental
characterization is very difficult. An interesting  approach, from
this point of view, is the use of atomic and molecular impurities
as probes \cite{sch90,goy92}. Theoretical calculations for
droplets with impurities have been recently done with both
Monte Carlo \cite{bar92} and density functional methods
\cite{dal89,dal93}. Impurities heavier than helium atoms are
expected to have small zero point motion, so that they can
be treated as classical objects in a quantum fluid
\cite{dal93}. This is certainly not true for   \het and hydrogen
impurities, for which a full quantum mechanical
treatment is needed. The case of \het impurities is particularly
interesting from the theoretical viewpoint. Since \het and \hef
atoms interact through the same potential, the properties of their
mixed systems are determined only by  quantum effects,
i.e., the different statistics and the different zero point
motion. One \het atom, being lighter than \hef, tends to move in
regions of low \hef density. This is the origin of the so called
Andreev state of \het on a \hef liquid-vapour interface at low
temperature  \cite{and66,saa71,edw78,man79,dal88,pav91a}. The same
behavior is observed in \hef films on solid substrates
\cite{ali92},  where the layer structure of the \hef density
produces a rich variety of \het states
\cite{she85,kro88,pav91,tre93}. Predictions on the Andreev states
of \het atoms on \hef droplets have been already given
\cite{dal89} using a phenomenological density functional, as well
as the variational Feynman-Lekner theory. A key
ingredient in the latter approach is the \hef kinetic energy
density, for which a simple approximate expression was
proposed in Ref. \cite{man79}.

In the present work  we calculate ground state properties of pure
\hef droplets by means of a variational Monte Carlo (VMC) method.
We follow the same procedure as in Ref. \cite{pan86}, but with a
different parametrization for the variational wave function.
A detailed discussion about the results for the ground state of
the droplets is given  elsewhere \cite{newvmc}. Here we present
the first microscopic results for the kinetic energy density,
which are shown to be in good agreement with the predictions of
Ref. \cite{man79}. Finally,  we use the  ground state properties
of \hef droplets as an input in the Feynman-Lekner theory for
\het impurities. We calculate the binding energy of the \het
surface states for several \hef droplets. The results are
extrapolated to estimate the binding  energy of \het atoms on a
planar \hef surface, in good agreement with the experimental
value \cite{edw78}.

The work is organized as follows. Section II contains the
short description of the VMC method and the results for the
kinetic energy density of \hef droplets. The Feynman-Lekner
approach is briefly introduced in Section III, where the results
for the Andreev states are also discussed. Section IV is devoted
to conclusions.

\section{Particle and kinetic energy density of pure \hef droplets}

First, we want to calculate the ground state energy, density
profile, and kinetic energy density for droplets of given
number of particle $N$. We use the VMC
method. In the variational approach a suitable form
for the many body  wave function $\Psi_v ({\bf r}_1, \dots ,{\bf r}_N)$
is choosen,
containing  a set of parameters that are varied in order to
minimize the energy of the system. The  Hamiltonian has the
usual form
\begin{equation}
H_0= \sum_{j} { - \hbar^2  \over 2 m_4 } \nabla_j^2 +
\sum_{i < j} v (r_{ij}) \ \ \ ,
\label{h0}
\end{equation}
where $m_4$ is the mass of \hef atoms. Aziz HFDHE2 potential
\cite{azi} is used as the pair interatomic  potential $v(r_{ij})$.

Several forms of the variational wave function of the pure \hef liquid
have been proposed in the literature \cite{pan83,pan86,wha90,kro90,newvmc}.
They are expected to give very similar results in the context of the
present work.  We use the  formalism of Ref. \cite{newvmc}, where
the wave function is taken of the same form as in Ref. \cite{pan86},
\begin{equation}
\Psi_v = \prod_{i\leq N} f_1(r_i) \prod_{i<j\leq N} f_2 (r_{ij})
\prod_{i<j<k \leq N} f_3 (r_{ij},r_{jk},r_{ki}) \ \ \ ,
\label{psi}
\end{equation}
but with an improved version of the correlation functions $f_1$,
$f_2$, and $f_3$. This choice ensures a correct asymptotic
behavior of the wave function far outside the droplet, which
has an important role for the \het and
hydrogen  impurity states.

Once the optimal  wave function is found, it is used to calculate
the one-body density
and the kinetic energy density, given by
\begin{equation}
\rho ({\bf r}_1) = N \int d {\bf r}_2 \dots d {\bf r}_N |\Psi_v|^2 \ \ \ ,
\label{density}
\end{equation}
and
\begin{equation}
\tau ({\bf r}_1) = {N \over \rho ({\bf r}_1)} \int d{\bf r}_2
\dots {\bf r}_N \Psi_v^* \left( {-\hbar^2 \over 2 m_4 } \nabla_1^2
\right) \Psi_v \ \ \ ,
\label{tau}
\end{equation}
respectively.

We consider droplets with 70,
112, 168, 240, 330, and 728 particles. The resulting density
profiles and the kinetic energy density  are shown in
Fig.\ \ref{fig1}.  It is worth noticing that both Monte Carlo
\cite{pan83,pan86,cep89,wha90,kro90,newvmc}
and  density functional \cite{str87,dal93} calculations predict an
extrapolated free surface thickness (distance between the points
where the density is 90\% and 10\% of the bulk value) of the order
of  $6 \div 8$ \AA, which is consistent with our results.

The kinetic energy density is shown in the lower part of
Fig.\ \ref{fig1}. It decreases smoothly from the inner value, close
to the value of the kinetic energy per particle in bulk liquid, to
the asymptotic limit $\mu_4(N)$, i.e., the chemical potential of the drop
containing $N$ \hef atoms. This
limit follows from the behavior of the one-body factor which in our
parametrization \cite{newvmc} is
\begin{equation}
f_1 (r \to \infty) \propto \sqrt{\rho (r \to \infty)} \propto \ {1\over r}
\exp
\left[ -  \left( {2 m_4 | \mu_4 | \over \hbar^2 } \right)^{1\over2}
r    \right] \ \ \  ,
\label{psiinf}
\end{equation}
which dominates the $r \to \infty$ limit of the wave function, yielding
\begin{equation}
\tau(r \to \infty) = \mu_4 \ \ \ .
\label{tauinf}
\end{equation}

In Ref. \cite{man79} the  approximate expression
\begin{equation}
\tau = \tau_0 \left( {\rho \over \rho_0} \right)^n -
{ \hbar^2 \over 2 m_4} { \nabla^2 \sqrt{\rho} \over \sqrt{\rho} } \ \ \ ,
\label{tauapp}
\end{equation}
was proposed for $\tau$,  where $\tau_0$ and $\rho_0$ are the ground state
kinetic energy per particle and the particle density in bulk \hef at zero
pressure, while $n$ is a phenomenological parameter.
Equation (\ref{tauapp}) is an interpolation between the
expected behavior of $\tau$ in the two opposite limits  $\rho \to
\rho_0$  and $\rho \to 0$. It has never been  checked so far with
microscopic  calculation. In Ref. \cite{man79} the values of $\tau_0$
and $n$ were fixed in a phenomenological way to reproduce  known
properties of \het impurities in bulk liquid \hef, while the saturation
density was taken directly from experiments ($\tau_0=13.34$ K, $n=1.76$,
$\rho_0=0.365 \sigma^{-3}$, where $\sigma=2.556$ \AA).  The same
quantities can be calculated  microscopically. For instance, the kinetic
energy per particle in bulk liquid at several densities was calculated
in Ref. \cite{usm82} using the variational wave function (\ref{psi}).
{}From those data one extracts $\rho_0=0.362 \sigma^{-3}$, $\tau_0=14.52$ K
and $n=1.77$. In Fig. \ref{fig2} we compare the VMC results for $\tau$,
for droplets with 112 and 728 atoms, with the ones of the approximated
formula (\ref{tauapp}). The solid line corresponds to the variational
values of $\tau_0$, $n$ and  $\rho_0$, while the dashed line corresponds
to the phenomenological parameters of Ref. \cite{man79}. In both cases
we have used Eq. (\ref{tauapp}) with the VMC density of the corresponding
droplets, and the high frequency statistical fluctuations have been filtered
out  in the calculation of the second derivative. The figure reveals
that the approximate formula (\ref{tauapp}) works very well, especially
for large droplets.

\section{Lekner-Feynman theory for \het impurities}

Once the ground state properties of pure \hef droplets are
obtained, one can calculate the energy and wave function of one
\het impurity. Consider a droplet consisting of $(N-1)$ \hef atoms
and one \het impurity atom. Since  the interatomic potential  is
the same for \het and \hef, one  can write the Hamiltonian in
the form
\begin{equation}
H=H_0 + H_I  \ \ \  ,
\end{equation}
where $H_0$ is given in Eq. (\ref{h0}) and
\begin{equation}
H_I =  - {\hbar^2 \over 2 m_4 } \left( {m_4 \over m_3} -1 \right)
\nabla_1^2 \ \ \  .
\end{equation}
A realistic trial wave function would be
\begin{equation}
\Psi = f({\bf r}_1) \prod_{i=2}^N F({\bf r}_1,{\bf r}_i) \Psi_0 \
\ \ ,
\label{psiimp}
\end{equation}
where $\Psi_0$ is the ground state of the Hamiltonian $H_0$,
and $f$ and $F$ are variational functions to be determined. In
principle, the optimal $f$ and $F$ can be found by
minimizing the total energy of the system $\langle \Psi | H |
\Psi \rangle$, which can be done either by solving resulting Euler equations
or by direct Monte Carlo simulations.

A significant simplification is obtained by taking  $F({\bf r}_1,{\bf r}_i)
\equiv 1$
\cite{lek70}. This corresponds to the assumption that the correlations between
\het impurity and \hef atoms are the same as those among \hef atoms.
The Euler equation for $f$ then becomes
\begin{equation}
-{\hbar^2 \over 2 m_3} {\bf \nabla}_1 \cdot ( \rho({\bf r}_1) {\bf
\nabla_1} f({\bf r}_1) ) + \left( {m_4 \over m_3} -1 \right)
f({\bf r}_1) \tau({\bf r}_1) \rho({\bf r}_1) -
\lambda f({\bf r}_1) \rho({\bf r}_1) = 0 \ \ \  ,
\end{equation}
where $\rho$ and $\tau$ are the particle  and  kinetic
energy densities of the pure \hef droplet (see Eqs. (\ref{density})
and (\ref{tau})). The same equation can be rewritten in the form of
a Schr\"odinger  equation
\begin{equation}
-{\hbar^2 \over 2 m_3} \nabla^2 \chi ({\bf r}) + V_3 ({\bf
r}) \chi({\bf r}) = \epsilon \chi ({\bf r}) \ \ \ ,
\label{schhe3}
\end{equation}
where
\begin{equation}
\chi ({\bf r}) =  f({\bf r}) \sqrt{\rho({\bf r})}
\end{equation}
is the impurity wave function, and
\begin{equation}
V_3({\bf r}) = \left({m_4 \over m_3} -1 \right) \tau({\bf r}) +
{\hbar^2 \over 2 m_3} { \nabla^2 \sqrt{\rho({\bf r})} \over
\sqrt{\rho({\bf r})} } + \mu_4
\label{v3}
\end{equation}
is an effective potential seen by the \het atom. The chemical potential
$\mu_4$ of the droplet of the considered size  is added for convenience,
so that the eigenvalue $\epsilon$ in Eq. (\ref{schhe3}) is referred to
the state in vacuum.
In principle, an additional term should appear in the effective potential
$V_3({\bf r})$ because the exact ground state of pure \hef droplet $\Psi_0$ is
approximated by the variational wave function $\Psi_v$.
This term is expected to be small due to the closeness of these two wave
functions, and is neglected in the present treatment.

The assumption  $F({\bf r}_1,{\bf r}_i) \equiv 1$ in Eq. (\ref{psiimp})
corresponds to the
Feynman-Lekner theory \cite{lek70}, already used in the past to predict
the properties of the Andreev state on the \hef free surface
\cite{saa71,man79}, on films \cite{tre93}, as well as on
droplets \cite{dal89}. Unlike in the previous calculations, where
phenomenological $\rho$ and $\tau$ were used, here we take them
from {\it ab initio} calculations.

In order to find the energy and the wave function of the
impurity on a droplet Eq. (\ref{schhe3}) is rewritten in
spherical coordinates:
\begin{equation}
- {\hbar^2 \over 2 m_3} {d^2 \over dr^2} \chi_{nl} +
\left( V_3 + {\hbar^2 \over 2 m_3} {l(l+1) \over r^2 }
\right) \chi_{nl} = \epsilon_{nl} \chi_{nl} \ \ \ .
\end{equation}
The crucial point is that the effective potential $V_3$ has a well
on the surface of the droplet, so that the lowest eigenstates of
the Schr\"odinger equation are localized on the surface. The
potential well originates from a  balance between the excess
kinetic energy of an \het atom in the bulk with respect to the one of
\hef atoms, which tends to push the \het atom out, and the He-He
interaction, which binds the \het atom to the liquid. A typical
situation is shown in Fig. \ref{fig3} for a droplet of 112
atoms. Results for the energy of the lowest ($n=0$) impurity
states on six  droplets are given in Fig. \ref{fig4} as a
function of $N^{-1/3}$. We note that a suitable smoothing procedure
has been  applied  to the VMC data in order to avoid spurious effects
of statistical fluctuations of $\rho$  in the calculation of the second
derivative. This affects mainly the external tail of $V_3$
where the density is very small.  The error bars in Fig. \ref{fig4}
correspond to the consequent inaccuracy in the results, estimated  by
choosing different smoothing methods. The extrapolation to the case
of the planar surface ($N \to \infty$)  can be  done by a linear fit,
even if the accuracy of the fit is relatively poor. We obtain
$\epsilon(\infty)  \simeq -4.9$ K, rather close to the
experimental estimate  $\epsilon(\infty) = (-5.02 \pm 0.03)$ K
quoted in Ref. \cite{edw78}.

In Fig. \ref{fig4} the results of the VMC calculation are compared
with predictions of Ref. \cite{dal89} (full line) obtained using
the Feynman-Lekner theory with  approximation (\ref{tauapp}) for $\tau$
and density profiles of  Ref. \cite{str87}. The significant
difference between the two predictions is mainly due to the different
density profiles. The curvature of the surface profiles in
the outer region is underestimated in the density functional
calculations of Ref. \cite{str87} with respect to the VMC results (see
Fig. \ref{fig5}). This makes the potential well for the impurity
wider, and the binding energy lower. The density profiles calculated
with a more recent density functional \cite{dup90,dal93} are sharper than
the ones of Ref. \cite{str87} and the corresponding predictions for the
\het binding energy are closer to the VMC results (dashed lines in
Figs. \ref{fig4} and \ref{fig5}). Part of the remaining discrepancy
is due to the use of the approximated expression (\ref{tauapp})
for $\tau$; we have checked that the corresponding effect on the
\het binding energy is small (about  $0.1$ K).

By solving the Schr\"odinger equation (\ref{schhe3}) one finds,
above  the lowest eigenstate, a spectrum of states with
different principal quantum number $n$ and angular momentum $l$.
The general features of the spectrum are the same as in
Ref. \cite{dal89}. In the limit of an infinite droplet the
states with different $l$ coincide with those of a 2-dimensional
Fermi gas.

To conclude, we stress again the idea of the Feynman-Lekner
approach and its limits.  The idea is that the two-body correlations
between the \het impurity and \hef atoms are taken to be the same
as the ones between the \hef atoms. The \het wave function is
then expressed by means of the factor $f({\bf r})$ in the
many body wave function (\ref{psiimp}).
The form of $f$ is derived variationally, by solving a
Schr\"odinger-like equation. The lowest eigenvalues are localized
on the surface of the droplets, as an effect of the different mass
of \het and \hef.   To improve the Feynman-Lekner theory
one should account for the fact that \het-\hef correlations differ
from the \hef-\hef ones, i.e., one should take  $F({\bf r}_1,{\bf r}_i)\neq 1$.
One possibility is to choose  $F=(1+\phi)$, where $\phi$
is small. Keeping the terms in the expansion of the total energy
up to the second order
in $\phi$ is equivalent to taking into account the energy of an elastic
deformation of the residual droplet due to the presence of the
impurity. Such corrections have already been studied in Refs.
\cite{guy82} and  \cite{kro88} in the case of liquid
\hef films. The problem of \het-\hef correlations, in the context
of variational calculations, has been also discussed in Ref.
\cite{bor93,kro93} for bulk \het-\hef mixtures. Work in this
direction is in progress.

\section{Conclusions}

We present the variational Monte Carlo (VMC) calculation for \hef
droplets with and without \het impurities. We have used the variational
wave function of Ref. \cite{pan86} modified as in Ref. \cite{newvmc}.
The results for the density and energy of six  droplets, from $N=70$
to $N=728$, are close to the predictions of previous calculations.
We have calculated the kinetic  energy density, for which only
approximate estimates have  been given so far. Our results show
that the analytic formula given in Ref. \cite{man79} for the
kinetic energy density (Eq. (7)) works very well, and can be safely used
in calculations involving the \hef surface.

In the second part of the work we apply the Feynman-Lekner
approach to study \het impurity states on \hef droplets.
The impurity wave function turns out to be  localized on the
surface of the droplets, as expected. The lowest value of the binding energy
is almost linearly dependent on $N^{-1/3}$ and the
value extrapolated to $N \to \infty$ is close to the experimental
binding energy of the Andreev state on a planar surface.
We have also compared our results with the predictions for
the \het binding energies obtained using phenomenological
calculations of the density profiles and the kinetic energy density.
The general trend of these earlier results is similar to the one of VMC
calculations.




\begin{figure}
\caption{ Particle density and kinetic energy density for
droplets of 70 (squares), 112 (plus signs), 168 (circles),
240 (crosses), 330 (triangles), and 728 (diamonds) atoms. }
\label{fig1}
\end{figure}

\begin{figure}
\caption{ Kinetic energy density for droplets of 728 and 112
atoms. Circles: VMC; solid lines: Eq.(7) with $\rho_0=0.362
\sigma^{-3}$, $\tau_0=14.52$ K and  $n=1.77$; dashed lines: the same
equation  with $\rho_0=0.365 \sigma^{-3}$, $\tau_0=13.34$ K and
$n=1.76$. }
\label{fig2}
\end{figure}

\begin{figure}
\caption{ Lowest eigenstate of one \het impurity on a droplet of
112 atoms. Solid line: \het wave function; squares: \hef density;
circles: effective potential $V_3$. }
\label{fig3}
\end{figure}

\begin{figure}
\caption{ Binding energy of the lowest impurity state as a
function of $N^{-1/3}$. Points with error bars: Feynman-Lekner
theory with $\tau$ and $\rho$ from VMC calculations;
solid line: Feynman-Lekner results of Ref.~[10]; dashed line:
same theory but with more recent density functional calculations
for the density profiles, as in Ref.[11].}
\label{fig4}
\end{figure}

\begin{figure}
\caption{ Surface profile for a droplet of 112 atoms. Points: VMC
results; solid line: density functional results of Ref.~[3]; dashed
line: density functional results of Ref.[11]. }
\label{fig5}
\end{figure}


\end{document}